# The Design of a Mobile App for Promotion of Physical Activity and Self-Management in Prostate Cancer Survivors: Personas, Feature Ideation and Low-Fidelity Prototyping

Francisco Monteiro-Guerra
Salumedia Tecnologías
Seville, Spain
University College Dublin
Dublin, Ireland
franciscoguerra@salumedia.com

Octavio Rivera-Romero
Universidad de Sevilla
Seville, Spain
Salumedia Tecnologías
Seville, Spain
orivera@us.es

Vasiliki Mylonopoulou
University of Oulu
Oulu, Finland
vasiliki.mylo@oulu.fi

Gabriel R. Signorelli
Oncoavanze
Seville, Spain
University College Dublin
Dublin, Ireland
g.signorelli@oncoavanze.es

Francisco Zambrana
University College London Cancer Institute
London, United Kingdom
Spanish Oncology Genitourinary Group
Madrid, Spain
f.zambrana@ucl.ac.uk

Luis Fernandez-Luque
Qatar Computing Research Institute
Doha, Qatar
Salumedia Tecnologías
Seville, Spain
lluque@qf.org.qa

*Abstract*—Most prostate cancer survivors are confronted with disease-related and treatment-related side effects that impact their quality of life. A tool that combines specific physical activity coaching with the promotion of a healthy lifestyle and self-management guidance might be a successful method to enhance a lifestyle change in these patients. As a prerequisite for useful health technology, it is important to consider a design process centred in the patients.
The aim of this study was to investigate the context of the problem and the user needs to support the ideation of a low-fidelity prototype of a tool to promote a healthy lifestyle among early-stage prostate cancer survivors.
A user-centred design approach was followed involving a multidisciplinary team. The prototype was developed in 3 phases. In phase 1, the context was studied with 2 systematic reviews of the state of practice and consulting with 3 specialists in Oncology, resulting in a global use case and main requirements. In phase 2, the needs and barriers of the users were studied based on literature research and validated with 3 specialists, resulting in the creation of 3 personas. In phase 3, 2 sessions were held to ideate and prioritize possible app features, based on brainstorming and selection techniques. Using the Ninja Mock and Proto.io software a low-fidelity prototype was developed, resulting in 25 interactive screens.
Understanding the user needs and context seems to be essential to highlight key goals, hence facilitating the bridge between ideation of the tool and the intended users' tasks and experiences. The conclusion of this first stage of the design process brings valuable details (such as barriers of the users to technology and to physical activity) for future iterations of design of the mobile app.

## I. INTRODUCTION

Prostate cancer (PCa) is estimated to be the most frequently diagnosed cancer in men in 2017, and the third cause of cancer-related deaths among men [1]. Advancements in PCa therapies have improved cure-rates and survival [2], making post-treatment quality of life a key issue. Both disease-related symptoms and treatment-related side effects may impact on the well-being and the physical, social and emotional functioning of these patients. For instance, radical prostatectomy and radiotherapy may cause adverse effects such as erectile dysfunction, urinary incontinence, and bowel problems [3]. Hormonal manipulations intended to decrease testosterone levels may lead to muscle wasting, hot flushings, decrease bone density and metabolic abnormalities that may even increase mortality [4]. Hence, PCa-specific health-related quality of life also refers to general well-being, vitality, fatigue, pain, general health status, global quality of life, and life satisfaction [5][6][7]. Unfortunately, physicians often underestimate these symptoms and its impact during treatment and on the long-term after primary treatment modality.

Between the general recommendations for prevention of cancer recurrence and improvement of the quality of life of patients is the practice of physical activity [8]. Rapidly accumulating research is demonstrating that routine exercise throughout and after treatment can provide important benefits including, among others, improving self-esteem, emotional stability, fatigue, and exercise capacity in men with PCa [9][10][11]. Despite the growing evidence, exercise is infrequently discussed between practitioners and patients and it prevails a lack of information about the available resources for exercise, risk level, contraindications, and correct prescription of exercise for these patients [12]. Cancer exercise programs delivered by certified exercise physiologists with cancer-specific training and supervised by medical staff are becoming more prevalent but still represent the exception in most oncology clinics [12].

Therefore, PCa survivors require support in developing skills to be in charge of their own health, enabling motivation to change. To accomplish the need of supporting these patients in self-management of their health burdens and to engage with physical activity, new tools are required.

Even with a considerable increase in technological solutions for cancer patients, the amount of mobile solutions developed for self-care and management of the disease is low [13]. The goal of the project is to develop a mobile app for health promotion that will address the gap in supportive tools for PCa survivors. The design of the app will focus on the promotion of physical activity, while providing the necessary support and formation tolls for self-management of disease and treatment side effects. The overall objective will be to empower PCa survivors to engage in a positive lifestyle change.

As a prerequisite for useful technology and a successful intervention that meets the requirements and preferences of the patients, it is important to follow a user-centred design (UCD) process [14]. UCD is a multi-stage, problem-solving process that investigates the needs, desires and limitations of users to increase the success rate of the usability in computerized systems [15]. This paper reports on a first iteration of such design process, which led to development of a first low-fidelity prototype of the proposed mobile app. The intention of this study will be to address the needs and barriers of prostate cancer survivors, and its implications in the development of a tool for health promotion in this population.

## II. METHODS

The stages completed in this study are part of a UCD approach for the development of the proposed app. The phases proposed for the development of this app are based on a study for the development of a mobile tool to stimulate physical activity of people with chronic diseases in primary care [16].

To understand the context of use and the user requirements, personas (representative fictional characters) were created as part of the design methodology [17]. Despite the broadly defined UCD flow in which a persona is ideally created after synthesis of user studies, design is not a linear process [18] and it is also believed that a persona can be born at other stages in the design process [18]. This is especially important when the end-users are patients, as their involvement needs to be carefully planned to avoid overload. Therefore, in this first iteration of the project it was decided to gather user specific information from medical specialists and literature review.

A multidisciplinary team guided the study and included 4 experts from different backgrounds: Computer Science, Biomedical Engineer, Human Computer Interaction and Sports Sciences.

The first iteration of the UCD process followed three iterative stages, depicted in Fig. 1.

*Phase 1: Context Understanding*

   *a) State of the Practice*

Two reviews were performed in March 2016 to address apps available in the literature and market.

A literature review was conducted to gather information about apps for PCa patients reported in scientific publications. A combination of medical and technological keywords was used in the searches: "prostate cancer", "app", "apps", "mobile", and "mhealth". The searches were performed on MEDLINE database with the use of selection criteria. Then a title and abstract review was performed.

An app review following a similar methodology used in [19][20][21] was conducted. "Prostat* cancer" was used as the searching string. The search was performed both in Google Play and Apple Store with the same selection criteria. Then, a review of the title and description of apps were conducted.

   *b) Global Use Case*

Literature on the topic of physical activity (e.g.: [22][23][24]) and self-management for PCa survivors was studied and discussed with 1 Oncologist and 2 Psycho-oncologists resulting in the definition of a global use case and the main content requirements of the app. The global use case was designed based on the methodology used in [16], to describe the interaction between a user and the system. The use case facilitates the communication of the conceptual idea of the tool to the design team, without giving too much direction to their thoughts.

*Phase 2: User Understanding*

   *a) Target Population*

Based on background literature the end-users of the tool were identified. This resulted in the definition of inclusion criteria that took into consideration the objective defined in the use case and the different subgroups of the PCa population.

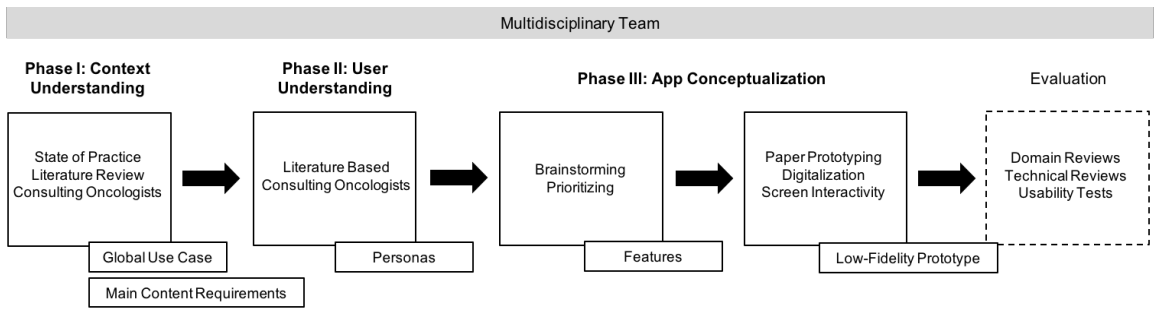

**Figure 1: Schematic of the study first UCD iteration.** First, the context was understood based on the state of practice and literature review, resulting in the global use case and main content requirements, defined together with oncologists. Second, the user needs were understood based on literature and consulting with oncologists, resulting in personas. In the third phase started the conceptualization of the tool, based on feature ideation, brainstorming and prototyping, resulting in a low-fidelity prototype.

*b) Personas*

The elaboration of personas was based on the methodology from [25] and [17]. Information was gathered from literature [3][6] as well as ideation from the team [18].

Taking into consideration the target population, users were categorized based on personal, disease-related factors, likes, fears, lifestyle and needs/goals.

The purpose is to understand, describe, focus and clarify user's goals and behaviour patterns.

*c) Validation of Personas*

The developed personas were validated based on domain review by two oncologists and one psycho-oncologist, in which the personas were presented in a narrative way, resulting in some minor corrections and considerations.

*Phase 3: App Conceptualization*

*a) Idea Generation*

To elicit features for the mobile app, 2 group sessions with the team were required, one for brainstorming and another for prioritizing and selecting.

A brainstorming technique was implemented based on the needs of the personas and the main requirements from the global use case. The team created a list of features using the Affinity Wall method [26]. The main content requirements were used as categories and, for 10 minutes, the team members individually wrote their ideas for each of the categories. Once finished, similar ideas were combined and 5 ideas were selected for each category based on team agreement. The resulting list was then hierarchically ordered for each of the categories based on a Selection-List technique [25]. Based on the priority given and the team agreement, the main features were selected.

*b) Prototyping*

Based on the features generated, the team started a phase of paper prototyping. This phase consisted in the creation of a low-fidelity prototype [27]. First, the ideas were sketched on paper using generic smartphone templates, making sure to include all necessary buttons and annotating the instructions for screen interaction. Priority was not given neither to content nor visuals. In a second step, a digital version of the sketch was created using NinjaMock software. In the third and last step, the created screens were transferred to Proto.io software where the interactions with buttons and screens were simulated.

### III. RESULTS

*Phase 1: Context Understanding*

*a) State of the Practice*

From the review of PCa apps in scientific literature, 195 results were found, after removing duplicates, and only 9 met the inclusion and exclusion criteria. 2 of them were focused on self-management and follow-ups. 5 of them analysed the facilitators and barriers of adoption of mhealth solutions. Finally, one app consisted on a simulation of sound of running water as a support tool for patients with urinary problems.

A total of 164 apps were found and only 7 met the defined criteria. 6 of them focused on self-management information and only one focused on physical activity. The last one consisted on a list of exercises without evidences. In Apple Store, only 2 apps met the inclusion and exclusion criteria. Both focused on information about PCa for patients.

Although mobile apps have been reported having positive impact on self-management and promoting physical activity among patients, from these reviews it can be concluded that there is still a gap in the development of these apps for PCa patients.

Based on these findings, it was decided to plan a solution for the specified gap. The objective was traced by first defining a global use case and the main content requirements, described in the following section.

*b) Global Use Case:*

The resulting concept of the tool is presented here as a narrative scenario:

"Tom is a survivor of prostate cancer dealing with post-treatment health burdens. Tom is now using the mobile app following the advice of his Urologist. Tom will get information from the app to manage his health with the use of personalized motivational messages, support tools and specific advice. The mobile app will also provide the necessary to motivate and coach him to practice physical activity, with available resources, training plans, goals, reminders and monitoring."

The main content requirements were defined as:
- Self-management of disease-related and treatment-related side effects;
- Physical activity and sports;
- Healthy lifestyle;
- Psychological support and motivation.

*Phase 2: User Understanding*

*a) Target Population*

Different groups of the PCa population have diverse physiological and psychological states and therefore different support needs. Aiming at a restricted audience facilitates the development of the app in terms of design, engagement strategies, goals and in terms of physical activity coaching. Considering the objective of the intervention, the inclusion criteria was defined as: survivors of PCa type I and II with a level of digital literacy of at least some experience with smartphone interactions. A compatible age group was defined: between 50 and 65 years old.

The chosen age group does not represent the peak of incidence of PCa but represents when the incidence rises sharply [28]. Targeting this younger age group means a lower concern with age and advanced-disease related burdens and risks.

Considering the defined target population, personas were created to represent the different user types that might use the app. Users' background, demographics, psychographics, health challenges, needs, goals and behaviours were synthetized from scientific documentation on PCa.

*b) Personas*

Taking into consideration the spectrum of key needs and requirements of the target population, it was decided to create 3 personas to represent different possible subgroups of end-users.

One example of persona is presented in Fig. 2.

*c) Validation of Personas*

From the domain reviews, the personas suffered slight changes to adjust the profile to the most common patients seen by the oncologists. The changes were mainly regarding the agreement between cancer types, treatment and the consequent side effects. Also, the oncologists underlined several important factors:

- Patients are not comfortable sharing their sexual and urinary problems;
- These patients normally do not care or understand about having a healthy lifestyle;
- In a lot of cases, their partners are the ones active in the search for information;
- Some patients are afraid of exercise risks due to their low physical condition;
- Patients that suffer from incontinence are not comfortable with physical activity;
- Patients are normally stressed and irritated with their loss of quality of life;
- Family and friends have an important role in the change of behaviour of these patients.

*Phase 3: App Conceptualization*

*a) Idea Generation*

The brainstorming phase was based on the main content requirements and personas. In table 1 the main features resulting from a selection of the high priority ideas are presented.

*b) Prototyping*

The purpose of this stage was to further specify the conceptual idea. Based on the previous ideated features a

TABLE 1. APP FEATURES LIST

| Content Requirements | Features |
|---|---|
| Self-Management | Documentation with advice on management of disease and treatment side effects |
| | Links to external sources of information |
| Physical Activity | Prostate cancer-specific exercise plan; combination of resistance and aerobic training |
| | Daily activity monitoring (registered by patient) |
| | Daily fitness challenges (default from app and/or set by family and friends) |
| | Access to previous activity |
| | Library with categorized exercises to improve physical condition, with videos and detailed explanation (including Pelvic exercises) |
| Healthy Lifestyle | Documentation with advice on nutrition, relaxation and fitness |
| Psychological Support and Motivation | Motivational messages and notifications |
| | Community: profile; section for discussions; blog with motivational experiences from PCa survivors |
| | Monitoring pain and emotional state* |
| | Option to share achievements with family and friends* |

*. Features that are still not included in the prototype

first low-fidelity prototype was sketched, resulting in 25 screens. Following the description in [27], the paper prototype developed is low in visuals, low in content and navigation hierarchy, and high in interactivity.

The visual attributes are captured but do not represent the look of a live system. Also, the prototype only includes a summary of the content. To facilitate future use for design evaluation and idea communication, links in the prototype were created to allow interaction with the different screens created.

For the resulting design of the main screens see Fig. 3.

This first low-fidelity prototype will be used in the following iterations of the design process to perform the first usability tests with users and technical evaluations with technical experts.

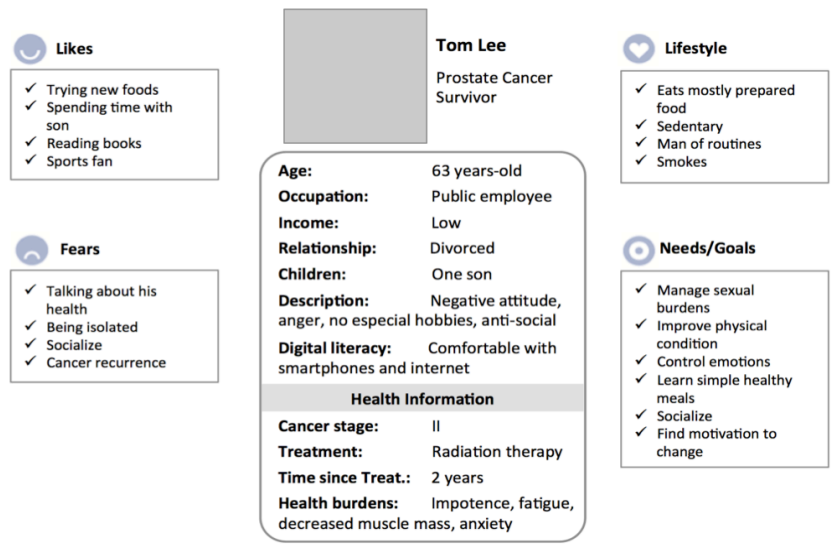

**Figure 2: An example of a prostate cancer survivor persona.**

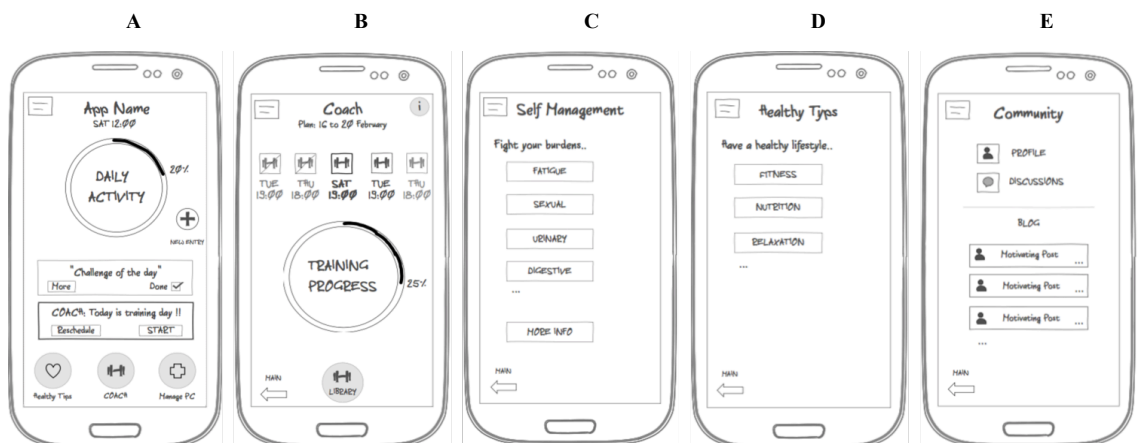

Figure 3: The main screens of the prototype interface. Screen A: includes important information of the current day as well as access for information on healthy tips and self-management; it shows the daily physical activity with the progress bar, allowing users to register any activity they take during the day with the 'New Entry' button; clicking the progress bar will take the user to a page with previous activity; the screen also includes one box with a daily challenge and another with a training reminder. Screen B: the 'Coach' screen with the proposed plan of training and access to the library of exercises; if the user clicks the icons over the schedule, it will take him to a training screen that starts with an explanation of the routine of exercises; progress in the overall training plan is shown with the progress bar. Screen C: the 'Self-Management' screen includes the documentation for managing the different possible health burdens. Screen D: the 'Healthy Tips' screen includes documentation with information on healthy lifestyle. Screen E: the 'Community' screen includes the option to create a profile, a discussion section and section for motivating blog posts.

## IV. DISCUSSION

To our knowledge this is the first study addressing the design of a mobile app for promotion of healthy lifestyle and self-management of PCa survivors. The first iteration of a UCD process was reported and proved to be valuable for the development of a first prototype of the proposed mobile app.

Design decision-making can be difficult and sometimes arbitrary [29]. The creation of personas helped this process by raising the team's awareness that there is a spectrum of key needs, desires, capabilities and limitations of the intended users. This was especially important, as the development team was not familiar with the end-users' context.

The success of this first design stage will shape the future design and development of the app. Personas have already proved to be valuable for a second iteration of design by bringing important information for the creation of a questionnaire for patients. Another possible outcome from the analysis of personas is the generation of scenarios for future testing and evaluation of the tool [30]. Also, these profiles can be used as the basis for selecting and screening users to participate in interactive prototyping activities and usability testing [29].

The development of the first low-fidelity prototype will bring value to the rest of the design process, as the idea of the tool can be effectively communicated with those involved and will allow for the first technical and users' evaluation. Working at this level of fidelity: facilitates changes; designers fell less wedded to a sketchy design; and users feel less pressure, being more likely to express reliable reactions [27].

At this early stage, credibility of the design was ensured with the involvement of healthcare specialists and using extensive literature investigation. Patients will be involved in future iterations to gather information for general app architecture, goal setting, feedback and data sharing [16].

We believe more people from different healthcare disciplines should have been involved in the validation of personas and brainstorming phase, to counteract the lack of patient input at this stage. However, the steps of design are time consuming and it is hard to involve these professionals throughout the process. To avoid overload, the involvement of both users and healthcare experts must be carefully planned for future design steps of the app.

To develop an effective intervention tool, it must be clear which determinants are relevant for the target behaviour and which of them can be influenced. This could be integrated by investigating motivation theories applied in similar tools and specific for physical activity and cancer survivorship [9]. In that context, engagement strategies such as quantified self, personalization, motivational messages and gamification might be further explored to integrate in the mobile app.

## V. FUTURE WORK

The process of UCD is an on-going development. New input from users and those interacting with them will be integrated in future iterations of the process.

The upcoming steps will involve: further evaluation of personas with oncologists, psycho-oncologists and patients; specify further requirements and preferences with focus groups; technical reviews of the prototype; and usability tests. Evaluation of the interface design, interaction and accessibility must be carefully taken into consideration [31]. This is especially important for the target users due to age limitations, as in digital literacy and poor vision. In a matured stage of the design, the team will address technical topics such as: architecture, wireframes; graphical user interfaces; and implementation in android studio.

## VI. Conclusion

In this study, a first low-fidelity prototype of a mobile app was developed with the involvement of a multidiscipline team and oncologists. Understanding the user needs and context was essential to highlight key goals, hence facilitating the bridge between ideation of the tool and the intended users' tasks and experiences.

The conclusion of this first stage of the design process will be important for future iterations of development of the mobile app. Based on current evidence in literature, it seems fair to assume that the success of such app might provide a solution for the current gap in supportive tools for prostate cancer survivors.

## Acknowledgements

This study has received funding from the European Union's Horizon 2020 research and innovation programme: Marie Sklodowska-Curie Actions grant agreement no. 722012 – CATCH - Cancer: Activating Technology for Connected Health; and Marie Sklodowska-Curie Actions grant agreement no. 676201. - CHESS - Connected Health Early Stage Researcher Support System. The work has been supported by Astellas through an unrestricted grant. Also, the study is part of the COST Action TD1405 European Network for the Joint Evaluation of Connected Health Technologies (ENJECT).

The study was performed in collaboration with Salumedia Tecnologias S.L., Astellas, the University of Seville – FIDETIA, OncoAvanze and the Spanish Oncology Genitourinary Group (SOGUG).

We would like to thank Pedro Valero Jiménez (MD) and Guido Giunti (MD, PhD).